\RequirePackage{fix-cm}
\documentclass[twocolumn,epjc3]{svjour3}  
\RequirePackage{graphicx}
\usepackage{cite} 
\usepackage{amsmath}
\usepackage{hyperref}
\usepackage{cite}
\usepackage{amsmath,amssymb}
\usepackage{color}

\begin{document}
\title{Ultracompact stars with polynomial complexity by Gravitational Decoupling}


\author{M. Carrasco--Hidalgo
\thanksref{        addr1} 
        \and
        E. Contreras\thanksref{e3,addr3} 

}
\thankstext{e3}{e-mail: 
\href{mailto:econtreras@usfq.edu.ec}{\nolinkurl{econtreras@usfq.edu.ec}}}

\institute{School of Physical Sciences and Nanotechnology, Yachay Tech University, 100119 Urcuqu\'i, Ecuador. \label{addr1}
\and
Departamento de F\'isica, Colegio de Ciencias e Ingenier\'ia, Universidad San Francisco de Quito,  Quito, Ecuador.\label{addr3}
}
\date{Received: date / Accepted: date}

\maketitle

\begin{abstract}
In this work we construct an ultracompact star configuration in the framework of Gravitational Decoupling by the Minimal Geometric Deformation approach. We use the complexity factor as a complementary condition to close the system of differential equations. It is shown that for a polynomial complexity the resulting solution can be matched with two different modified--vacuum geometries. 
\end{abstract}

\section{Introduction}\label{intro}

Through the developments of general relativity, black holes (BH) have been a subject of study, discussions and analysis. From being considered as simply mathematical constructions lacking of physical reality to be one of the central topics in recent researches, BH's are undoubtedly one of the most known and intriguing objects in literature. Nowadays, BH's are accepted as astrophysical objects and are considered as the preferable laboratory to test strong gravitational fields. Besides, recent observations of gravitational waves \cite{wv1,wv2,wv3} and black hole shadows \cite{ayima1,ayima2}
lead to the conclusion that BH indeed do exist. 

Although the existence of BH's seems undeniable, there are some aspects about this geometry which remains unclear. One of the main issues is the prediction of a space--time singularity which is often taken as an indicator that general relativity is incomplete and should
be generalized to a quantum theory in order to overcome this feature \cite{alfio}. In this direction, some efforts have been made  to provide a classical solution to this problem. All of these solutions are known as BH mimickers and 
correspond to objects that, given their high compactness, the motion of test particles around them seems indistinguishable when compared with the physics around a BH (see \cite{Cardoso:2019rvt}, for a recent review). Some examples of BH mimickers found in literature encompasses regular BH \cite{alfio,Bargueno:2020ais,Dymnikova:2020wve}, traversable wormholes \cite{Konoplya:2021hsm,Potashov:2020zmg,Bambi:2021qfo,blazquez} and ultracompact stars \cite{raposo}, among others, and in this work we shall focus our attention in the latter.

As it is well known, Buchdahl limit
relies on the hypothesis of isotropy
and entails that the maximum compactness of a self-gravitating, isotropic, spherically-symmetric object of mass $M$ and radius $R$ has an upper bound given by $M/R = 4/9$ (for modifications of the Buchdahl's limits induced by the presence of the cosmological constant see \cite{r1,r2,r3,r4}, for example ). In this regard, anisotropic self--gravitating fluids
enter as possibility to surpass such a limit and to provide well posed anisotropic ultracompact stars (for recent works in this direction, see Refs. \cite{raposo,ovalle2019a}, for example). An intriguing features of anisotropic ultracompact stars is that such models contain the Mazur and Mottola (MM) gravastar \cite{MM1,MM2} modeled as the
Schwarzschild interior in the ultracompact limit as a special case \cite{raposo,ovalle2019a}.

Recently, in Ref. \cite{ovalle2019a} the MM gravastar has been extended to anisotropic domains by
the well known gravitational decoupling (GD) \cite{ovalle2017} by the Minimal Geometric Deformation approach (MGD)  (for implementation in $3+1$ and $2+1$ dimensional spacetimes see
\cite{ovalle2014,ovalle2015,Ovalle:2016pwp,ovalle2018,ovalle2018bis,estrada2018,ovalle2018a,lasheras2018,estrada,rincon2018,ovalleplb,tello2019,lh2019,estrada2019,gabbanelli2019,sudipta2019,linares2019,leon2019,casadioyo,tello2019c,arias2020,abellan20,tello20,rincon20a,jorgeLibro,Abellan:2020dze,Ovalle:2020kpd,Contreras:2021yxe,Heras:2021xxz,contreras-kds,tello2021w,darocha1,darocha2,darocha3,sharif1,sharif2,sharif3,estrada2021}. The resulting model corresponds to an ultracompact anisotropic star surrounded by a MGD--modified vacuum which fulfil the main requirements to describe a stable model: it is regular at the origin and its density is positive and decrease monotonically from the center outwards \cite{ovalle2019a}. For these reasons, our main goal here is twofold: to obtain an alternative MGD--like gravastar model and to provide another MGD--modified vacuum as the exterior geometry of the compact configuration.

A key point in the implementation of MGD is to provide an auxiliary condition to obtain the so--called decoupling function. Some examples are the so called mimic constraint for the pressure and the density, regularity condition of the anisotropy function, barotropic equation of state, among others. However, in this work we use the recently introduced complexity factor for self--gravitating fluids \cite{complex1}. In particular, we propose a polynomial complexity factor which contains the gravastar model reported in \cite{ovalle2019a} as a special case.

This work is organized as follow. In the next section we review the main aspects on GD. In section \ref{complexityF} we introduce the concept of complexity factor. Section \ref{UCS} is devoted to revisit the ultracompact Schwarzschild star and
in section \ref{UCSMGD} we obtain the new anisotropic ultracompoact star by MGD. Finally, some comments and conclusions are in the last section.

\section{Gravitational decoupling}\label{GD}
In this section we review some aspects of GD by MGD (for more details, see \cite{ovalle2017}). Let us start with the Einstein field equations (EFE)
\begin{eqnarray}\label{EFE}
G_{\mu\nu}=R_{\mu\nu}-\frac{1}{2}g_{\mu\nu}R=\kappa T_{\mu\nu}
\end{eqnarray}
where
\begin{equation}\label{energy-momentum}
    T_{\mu\nu} = T^{(s)}_{\mu\nu} + \alpha\theta_{\mu\nu}\;.
\end{equation}
In the above equation $\kappa=8\pi$ \footnote{In this work we shall use $c=G=1$.}, $T^{(s)}_{\mu\nu}$ represents the matter content of a known solution of Einstein's field equations, namely the {\it seed} sector,  and $\theta_{\mu\nu}$ describes 
an extra source that is coupled by means of the parameter $\alpha$. Such a coupling is introduced in order to control the effect of 
$\theta_{\mu\nu}$ on $T^{(s)}_{\mu\nu}$. Since the Einstein tensor satisfies the Bianchi identities, the total energy--momentum tensor  satisfies
\begin{equation}\label{divergencia-cero-total}
    \nabla_{\mu} T^{\mu\nu} = 0\;.
\end{equation}
It is important to point out that, whenever 
$\nabla_\mu T^{\mu\nu(s)} = 0$, the following condition necessarily must be satisfied
\begin{equation}\label{divergencia-cero-theta}
    \nabla_\mu \theta^{\mu\nu} = 0\;,
\end{equation}
and as a consequence, there is no exchange of energy-momentum tensor between the seed solution and the extra source $\theta^{\mu\nu}$ (the interaction is purely gravitational).\\

From now on, let us consider a static, spherically symmetric space-time 
sourced by 
\begin{eqnarray}
T^{\mu(s)}_{\nu}&=&\textnormal{diag}(\rho^{(s)},-p_{r}^{(s)},-p^{(s)}_{\perp},-p^{(s)}_{\perp})\label{tmunu}\\
\theta^{\mu}_{\nu}&=&\textnormal{diag}(\theta_{0}^{0},\theta_{1}^{1},\theta_{2}^{2},\theta_{2}^{2})\label{thetamunu}
\end{eqnarray}
and a line element given by 
\begin{eqnarray}\label{metric}
ds^{2}=e^{\nu}dt^{2}-e^{\lambda}dr^{2}
-r^{2}(\theta^{2}+\sin^{2}\theta d\phi^{2}).
\end{eqnarray}
Replacing (\ref{tmunu}), (\ref{thetamunu})
and (\ref{metric}) in (\ref{EFE}) and (\ref{energy-momentum}), the EFE read
\begin{eqnarray}
    \kappa \rho &=& \frac{1}{r^2} +
        e^{-\lambda}\!\left(\frac{\lambda'}{r} - \frac{1}{r^2}\right)\!,\label{mgd05}\\
    \kappa p_{r} &=& -\frac{1}{r^2} +
        e^{-\lambda}\!\left(\frac{\nu'}{r} + \frac{1}{r^2}\right)\!,\label{mgd06}\\
    \kappa p_{\perp} &=& \frac{1}{4}e^{-\lambda}\!\left(2\nu'' + {\nu'}^2 - \lambda'\nu' + 2\frac{\nu'-\lambda'}{r} \right)\!\label{mgd07}
\end{eqnarray}
where we have defined
\begin{eqnarray}
    \rho &=& \rho^{(s)} + \alpha\theta_{0}^{0} \;,\label{mgd07a}\\
    p_{r} &=& p_r^{(s)}-\alpha\theta_{1}^{1}   \;,\label{mgd07b}\\
    p_{\perp} &=& p_{\perp}^{(s)} -\alpha\theta_{2}^{2} \;.\label{mgd07c}
\end{eqnarray}
Note that 
given the non-linearity of Einstein’s equations the decomposition (\ref{energy-momentum}) does not lead to two set of equations, one for each source involved. Nevertheless, contrary to the broadly belief, such a decoupling is possible in the context of MGD as we shall demonstrate in what follows.\\

Let us introduce a \textit{ geometric deformation} in the metric functions given by 
\begin{eqnarray}
    \nu\;\; &\longrightarrow &\;\; \xi + \alpha g,\label{comp-temporal} \\
    e^{-\lambda}\;\; &\longrightarrow &\;\; e^{-\mu} + \alpha f\;,\label{comp-radial}
\end{eqnarray}
where $\{f,g\}$ are the so-called decoupling functions and $\alpha$ is the same free parameter that ``controls'' the influence of $\theta_{\mu\nu}$ on $T_{\mu\nu}^{(s)}$. In this work we shall concentrate in the particular case $g = 0$ and $f \ne 0$. Now, replacing (\ref{comp-temporal}) and (\ref{comp-radial}) in the system (\ref{mgd05}-\ref{mgd07}), we are able to split the complete set of differential equations into two subsets: one describing a seed sector sourced by the conserved energy-momentum tensor, $T_{\mu\nu}^{(s)}$
    \begin{eqnarray}
    \kappa\rho^{(s)} &=& \frac{1}{r^2} +
        e^{-\mu}\!\left(\frac{\mu'}{r} - \frac{1}{r^2}\right)\!,\label{mgd13}
        \\
    \kappa p_r^{(s)} &=& -\frac{1}{r^2} +
        e^{-\mu}\!\left(\frac{\nu'}{r} + \frac{1}{r^2}\right)\!,\label{mgd14}
        \\
    \kappa p_{\perp}^{(s)} &=& \frac{1}{4}e^{-\mu} \!\! \left( 2\nu'' + {\nu'}^2 
    - \mu' \nu' + 2\frac{\nu'-\mu'}{r} \right) \!,\nonumber\\\label{mgd15}
\end{eqnarray}
and the other set corresponding to quasi-Einstein field equations sourced by $\theta_{\mu\nu}$
 \begin{eqnarray}
    \kappa\theta^0_0 &=& - \frac{f}{r^2} -\frac{f'}{r}\,,\label{mgd16}\\
    \kappa \theta^1_1 &=& -f
        \left(\frac{\nu'}{r} + \frac{1}{r^2}\right)\!,\label{mgd17}\\
    \kappa \theta^2_2 &=& -\frac{f}{4} \left( 2\nu'' + 
    {\nu'}^2 + 2\frac{\nu'}{r}\right)
    -\frac{f'}{4} \left( \nu' + \frac{2}{r} \right)\!.\nonumber\\\label{mgd18}
    \end{eqnarray}
As we have seen, the components of $\theta_{\mu\nu}$ satisfy the conservation equation $\nabla_\mu \theta^{\mu}_{\nu}=0$, namely
\begin{eqnarray}\label{consthe}
    \theta'^1_{1} 
     - \frac{\nu'}{2} (\theta^0_0 - \theta^1_1)      - \frac{2}{r}(\theta^2_2 - \theta^1_1)=0.
\end{eqnarray}

Hitherto, we have discussed the general aspects of GD by MGD without any specification of the system under study. However, if the system under consideration is the interior of some stellar configuration, the solution obtained will be valid only up to certain radius $R$ which define the surface of the star. In this regard, the matching between the interior solution with some exterior geometry for $r>R$ is mandatory and in most of the cases, it is sufficient to take
the Schwarzschild vacuum solution as the exterior metric. However, as it was demonstrated in \cite{ovalle2019a}, a suitable MGD--gravastar solution is possible whenever the exterior geometry is also a MGD--modified vacuum, namely
\begin{eqnarray}\label{sol04a}
    ds^2 = \left(1 - \frac{2M}{r}\right)dt^2 &-& \left(1 - \frac{2M}{r}+\alpha g(r)\right)^{-1} dr^2 +\nonumber\\ 
    &-& r^2 d\theta^2 - r^2\sin^2{\!\theta}\,d\phi^2\, ,\nonumber\\
\end{eqnarray}
with $g(r)$ is the decoupling function for the exterior solution. A list of MGD--modified vacuum solutions can be found in Ref. \cite{ovalle2018a}. Now, in order to match smoothly the interior metric with the outside one above on the boundary surface $\Sigma$, we require
\begin{eqnarray}
    e^{\nu}\Big|_{\Sigma^{-}} &=& \left(1 - \frac{2M}{r}\right)\Bigg|_{\Sigma^{+}}\,,\label{mgd11a}\\
    e^{\lambda}\Big|_{\Sigma^{-}} &=& \left(1 - \frac{2M}{r}+\alpha g(r)\right)^{-1}\Bigg|_{\Sigma^{+}}\,,\label{mgd11b}\\
     p_r(r) \Big|_{\Sigma^{-}} &=&      p_r(r) \Big|_{\Sigma^{+}}\,,\label{mgd11c}
\end{eqnarray}
which corresponds to the continuity of the first and second fundamental form across that surface. \\

To conclude this section, we would like to emphasize the importance of GD  by MGD as a useful tool to find solutions of EFE. As it is well known, in static and spherically symmetric spacetimes sourced by anisotropic fluids, EFE reduce to three equations given by  (\ref{mgd05}), (\ref{mgd06}) and (\ref{mgd07}) and five unknowns, namely $\{\nu,\lambda,\rho,p_{r},p_{\perp}\}$. In this sense, two auxiliary conditions must be provided: metric conditions, equations of state, etc. However, given that in the context of MGD a seed solution should be given, the number of degrees of freedom reduces to four and, as a consequence, only one extra condition is required. In general, this condition is implemented in the decoupling sector given by Eqs. (\ref{mgd16}), (\ref{mgd17}) and (\ref{mgd18}) as some equation of state which leads to a differential equation for the decoupling function $f$. In this work, we take an alternative route to find the decoupling function; namely, the complexity factor that we shall introduce in the next section.

\section{Complexity of compact sources}\label{complexityF}
Recently, a new definition for complexity for self--gravi--\nobreak tating
fluid distributions has been introduced in
Ref. \cite{complex1}. This definition is based on the intuitive idea that the least complex gravitational system should be characterized by a homogeneous energy density distribution and isotropic pressure. Now, as demonstrated in \cite{complex1}, there is a scalar associated to the orthogonal splitting of the Riemann tensor \cite{LH-C2} in spherically symmetric space--times which capture the essence of what we mean by complexity, namely
\begin{eqnarray} \label{YTF2}
Y_{TF} = 8\pi \Pi - \frac{4\pi}{r^3}\int^{r}_{0} \tilde{r}^3 \rho' d\tilde{r},
\end{eqnarray}
with $\Pi\equiv p_{r}-p_{\perp}$. Also, it can be shown that (\ref{YTF2}) allows to write the Tolman mass as,
\begin{eqnarray} \label{m_T}
m_{T} = (m_{T})_{\Sigma}\left(\frac{r}{r_{\Sigma}}\right)^3 + r^3\int^{r_{\Sigma}}_{r} \frac{e^{( \nu + \lambda )/2}}{{\tilde{r}}} Y_{TF} d\tilde{r},
\end{eqnarray}
which can be considered as a
solid argument to define the complexity factor by means of this scalar given that this function, encompasses all the modifications produced by the energy density inhomogeneity and the anisotropy of the pressure on the active gravitational mass. \\

Note that the vanishing complexity condition ($Y_{TF}=0$) can be satisfied
not only in the simplest case of isotropic and homogeneous system but in all the cases where
\begin{eqnarray}
\Pi=\frac{1}{2r^{3}}\int\limits_{0}^{r}\tilde{r}^{3}\rho' d\tilde{r}.
\end{eqnarray}
In this respect, the vanishing complexity condition leads to a non--local equation of state that can be used as a complementary condition to close the system of EFE (for a recent implementation, see \cite{casadioyo}, for example). Similarly, we can provide a particular values of $Y_{TF}$ and use this information to find a family of solutions with the same complexity factor. An example of how this can be achieved can be found in \cite{casadioyo}. In this work, we shall propose a suitable value for the complexity factor to find a new solution for an ultracompact star.

\section{Ultracompact Schwarzschild star}\label{UCS}
In this section we briefly review the Schwarschild interior in the ultracompact regime. As it is well known, the metric for this configuration reads
\begin{eqnarray}
ds^{2}=e^{\nu}dt^{2}-e^{\lambda}dr^{2}-r^{2}(d\theta^{2}+\sin\theta^{2}d\phi^{2}),
\end{eqnarray}
where
\begin{eqnarray}
e^{\nu}&=&\frac{1}{4}(3\sqrt{1-H^{2}R^{2}}-\sqrt{1-H^{2}r^{2}})^{2}\\
e^{-\lambda}&=&1-H^{2}r^{2},
\end{eqnarray}
with $M$ the mass, $R$ the radius of the star and
\begin{eqnarray}
H^{2}=\frac{2M}{R^{3}}.
\end{eqnarray}
The Schwarzschild interior is sourced by a perfect fluid with uniform density $\rho=\rho_{0}$ and a pressure given by
\begin{eqnarray}\label{p-G}
p=\rho_{0}\left(\frac{1-H^{2}r^{2}-\sqrt{1-H^{2}R^{2}}}{3\sqrt{1-H^{2}R^{2}}-\sqrt{1-H^{2}r^{2}}}\right).
\end{eqnarray}
At this point a couple of comments are in order. First, the Buchdahl limit set an upper bound of the compactness parameter, $M/R$, which entails a condition on the radius of the star, namely
\begin{eqnarray}
R>\frac{9}{4}M>2M.
\end{eqnarray}
The Buchdahl limit ensure that the pressure is finite and positive everywhere inside the star, as required for stable configurations. Second, note that the pressure (\ref{p-G}) is regular except at some radius $R_{0}$ given by
\begin{eqnarray}
R_{0}=3M\sqrt{1-\frac{8}{9}\frac{R}{M}}<R.
\end{eqnarray}
Now, as noted by Mazur and Mottola,
in the ultracompact limit, namely, 
when both $R$ and $R_{0}$ approach to the Schwarzschild radius $2M$, the interior solution corresponds to a patch of the
de Sitter solution. More precisely, the solution reads,
\begin{eqnarray}
e^{\nu}&=&\frac{1}{4}(1-H^{2}r^{2})\label{nuG}\\
e^{-\lambda}&=&1-H^{2}r^{2}
\label{lambdaG}\\
p&=&-\rho=\textnormal{constant}\label{eosG},
\end{eqnarray}
for $r<2M$. It can be shown that the above solution join
with the Schwarzschild vacuum in a way that the Israel second junction condition is violated. This  implies that the presence of a 
$\delta$--distribution of stresses is necessary to give a correct interpretation of the Schwarzschild star beyond the Buchdahl limit. However, as we shall demonstrate in what follows, the ultracompact solutions obtained by MGD does not require the existence of such a distribution of matter given that the metric functions join smoothly through the surface $\Sigma$.

\section{Ultracompact star by gravitational decoupling}\label{UCSMGD}
The MM gravastar
given by (\ref{nuG}), (\ref{lambdaG}) and (\ref{eosG})
was recently extended by MGD
in \cite{ovalle2019a}. In this case, the metric reads
\begin{eqnarray}
e^{\nu}&=&\frac{1}{4}(1-H^{2}r^{2})\label{nujorge}\\
e^{-\lambda}&=&1-H^{2}r^{2}+\alpha f(r)\label{lambdajorge},
\end{eqnarray}
with 
\begin{eqnarray}\label{fjorge}
f(r)=(1-H^{2}r^{2})H^{n}r^{n},
\end{eqnarray}
and $\alpha\ge-1$ to ensure that $g^{rr}$ is positive definite as $r\to2M$.
As demonstrated in \cite{ovalle2019a}, the above solution is ill--matched with the Schwarzschild vacuum because it requires $\alpha=-3/2<-1$ which violates the previous requirement for $\alpha$ to ensure the correct behaviour of $g^{rr}$. To overcome this difficulty, the proposed exterior solution was the modified vacuum 
\begin{eqnarray}
e^{\nu}&=&=1-\frac{2M}{r}\label{nuExt}\\
e^{-\lambda}&=&\left(1-\frac{2M}{r}\right)\left(1+\frac{\ell}{2r-3M}\right)\label{lambdaExt},
\end{eqnarray}
with $\ell$ a constant with units of a length. The reader is referred to Ref. 
\cite{ovalle2018a} where the MGD--modified vacuum (\ref{nuExt}) and (\ref{lambdaExt}) was obtained and discussed in detail.
It is worth mentioning that the decoupling function (\ref{fjorge}) leads to stable interior solutions only for $n=2$.\\

In order to provide an alternative MGD--gravastar solution, in this work we use the complexity factor previously introduced in Sect. \ref{complexityF}, as an auxiliary condition to close the system and find the decoupling function $f$. It can checked that, imposing the broadly used vanishing complexity condition, it does not lead to well behaved interiors so in this work we shall provide another value for the complexity and we shall use the solution given by (\ref{nujorge}), (\ref{lambdajorge}) and (\ref{fjorge}) as a guide. \\

A straightforward computation reveals that for $n=2$, the MGD gravastar model of Ref. \cite{ovalle2019a} has a complexity given by
\begin{eqnarray}\label{complexjorge}
Y_{TF}=\alpha H^{4}r^{2}.
\end{eqnarray}
Based on the above result, in this work we propose a polynomial complexity, namely
\begin{eqnarray}\label{complexpoly}
Y_{TF}=\sum^{N}\limits_{i=0}a_{i}r^{i},
\end{eqnarray}
which contains (\ref{complexjorge}) as a particular case. Indeed, (\ref{complexjorge}) is recovered for $N=2$, $a_{0}=a_{1}=0$ and $a_{2}=\alpha H^{4}$. Now, replacing (\ref{nujorge}) and (\ref{lambdajorge}) in 
(\ref{complexpoly}) we obtain
\begin{eqnarray}\label{complexf}
\frac{\alpha  H^2 r \left(H^2 r \left(2 f-r f'\right)+f'\right)}{2 \left(H^2 r^2-1\right)^2}=\sum^{N}\limits_{i=0}a_{i}r^{i},
\end{eqnarray}
which depending on the values of $N$, provides a differential equation for the decoupling function.\\

It can be shown that if either $a_{0}$ or $a_{1}$ are not vanishing factors, the solution of (\ref{complexf}) leads to divergent interior  
solutions so these possibilities must be discarded and, as a consequence,  (\ref{complexf}) now reads 
\begin{eqnarray}\label{complexfnmayor}
\frac{\alpha  H^2 r \left(H^2 r \left(2 f-r f'\right)+f'\right)}{2 \left(H^2 r^2-1\right)^2}=\sum^{N}\limits_{i\ge2}a_{i}r^{i},
\end{eqnarray}
which can be easily integrated to obtain
\begin{eqnarray}
f=-\frac{2}{\alpha H^{2}}(1-H^{2}r^{2})\sum\limits_{i\ge2}^{N}\frac{a_{i}r^{i}}{i}
\end{eqnarray}

For example, for $N=3$ we obtain \begin{eqnarray}\label{f-N3}
f=\frac{2}{\alpha  H^2} \left(1-H^2 r^2\right) \left(\frac{a_{2} r^2}{2}+\frac{a_{3} r^3}{3}\right),
\end{eqnarray}
from where
\begin{eqnarray}
e^{\nu}&=&\frac{1}{4}\left(1-H^{2}r^{2}\right)\\
e^{-\lambda}&=&\left(1-H^{2}r^{2}\right)\left(1-\frac{2}{H^{2}}
\left(\frac{a_{2}}{2}r^{2}+\frac{a_{3}}{3}r^{3}\right)\right)\\
\rho&=&-\frac{9 a_{2}+8 a_{3} r}{24 \pi  H^2}+\frac{15 a_{2} r^2+12 a_{3} r^3}{24 \pi }+\frac{3 H^2}{8 \pi }\\
p_{r}&=&\frac{3 a_{2}+2 a_{3} r}{24 \pi  H^2}-\frac{9 a_{2} r^2+6 a_{3} r^3}{24 \pi }-\frac{3 H^2}{8 \pi }\\
p_{\perp}&=&\frac{a_{2}+a_{3} r}{8 \pi  H^2}-\frac{5 a_{2} r^2+4 a_{3} r^3}{8 \pi }-\frac{3 H^2}{8 \pi }.
\end{eqnarray}
It is worth mentioning that in 
Eq. (\ref{f-N3}) we have discarded the term with the integration constant because it leads to divergence in the interior of the configuration.\\

To proceed with the analysis, we need to provide an exterior geometry and we shall explore
two different MGD--modified vacuum.
\subsection{Exterior 1}
Let us consider
\begin{eqnarray}
e^{\nu}&=&1-\frac{2M}{r}\\
e^{-\lambda}&=&\left(1-\frac{2M}{r}\right)\left(1+\frac{\ell}{2 r-3 M}\right)\\
\rho&=&-\frac{\ell M}{8 \pi  r^2 (3 M-2 r)^2}\\
p_{r}&=&-\frac{\ell}{24 \pi  M r^2-16 \pi  r^3}\\
p_{\perp}&=&\frac{\ell (M-r)}{8 \pi  r^2 (3 M-2 r)^2}
\end{eqnarray}
The continuity of the radial pressure requires
\begin{eqnarray}
a_{2}=-\frac{128 a_{3} M^6+3 l+9 M}{96 M^5},
\end{eqnarray}
while the continuity of the first fundamental form is automatically fulfilled.

\begin{figure}[h!]
\includegraphics[scale=0.5]{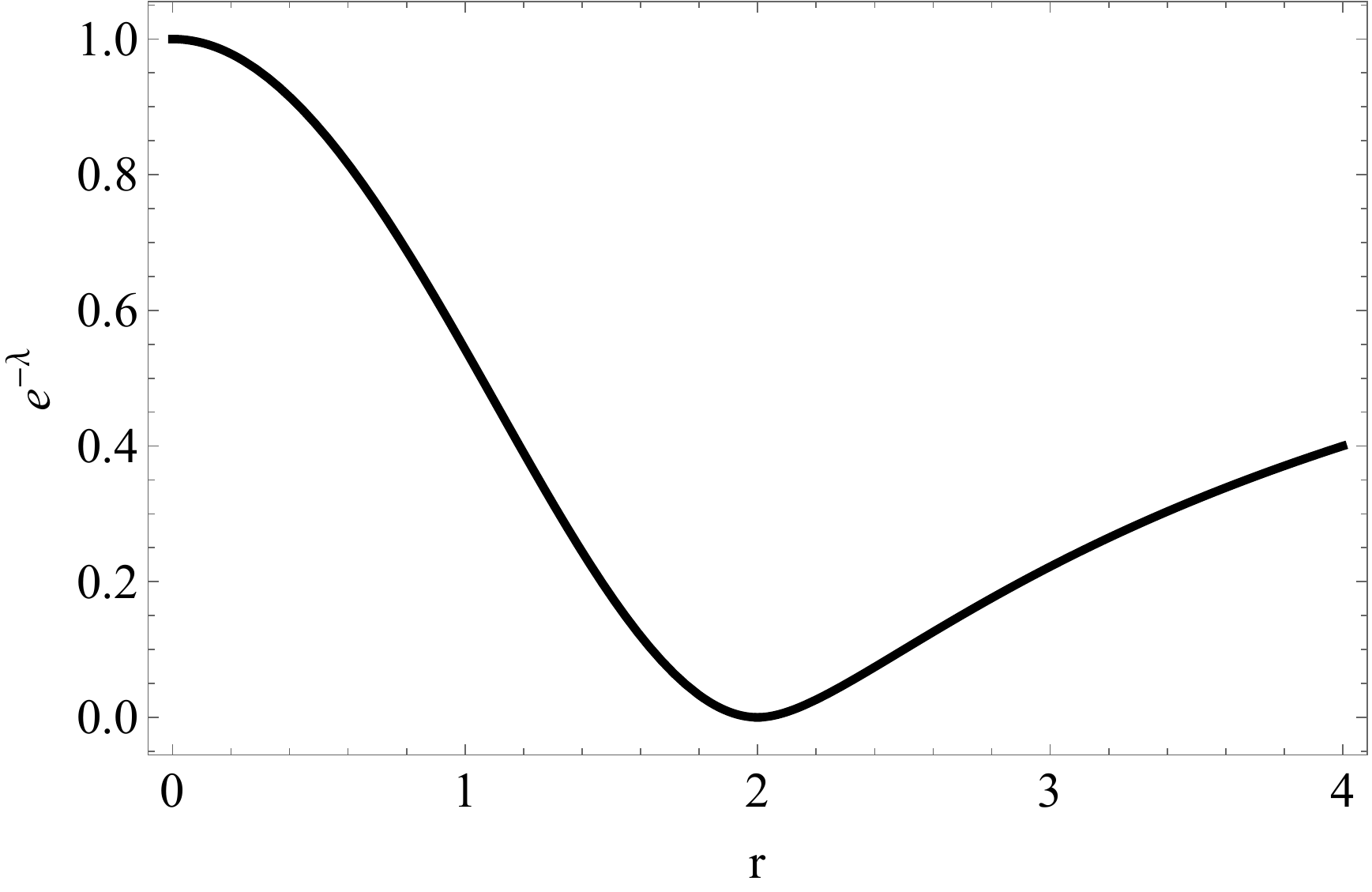}
\caption{\label{lambda} $e^{-\lambda}$ for $M=1$ $a_{3}=0.01$ and $l=-1$ }
\end{figure}

\begin{figure}[h!]
\includegraphics[scale=0.5]{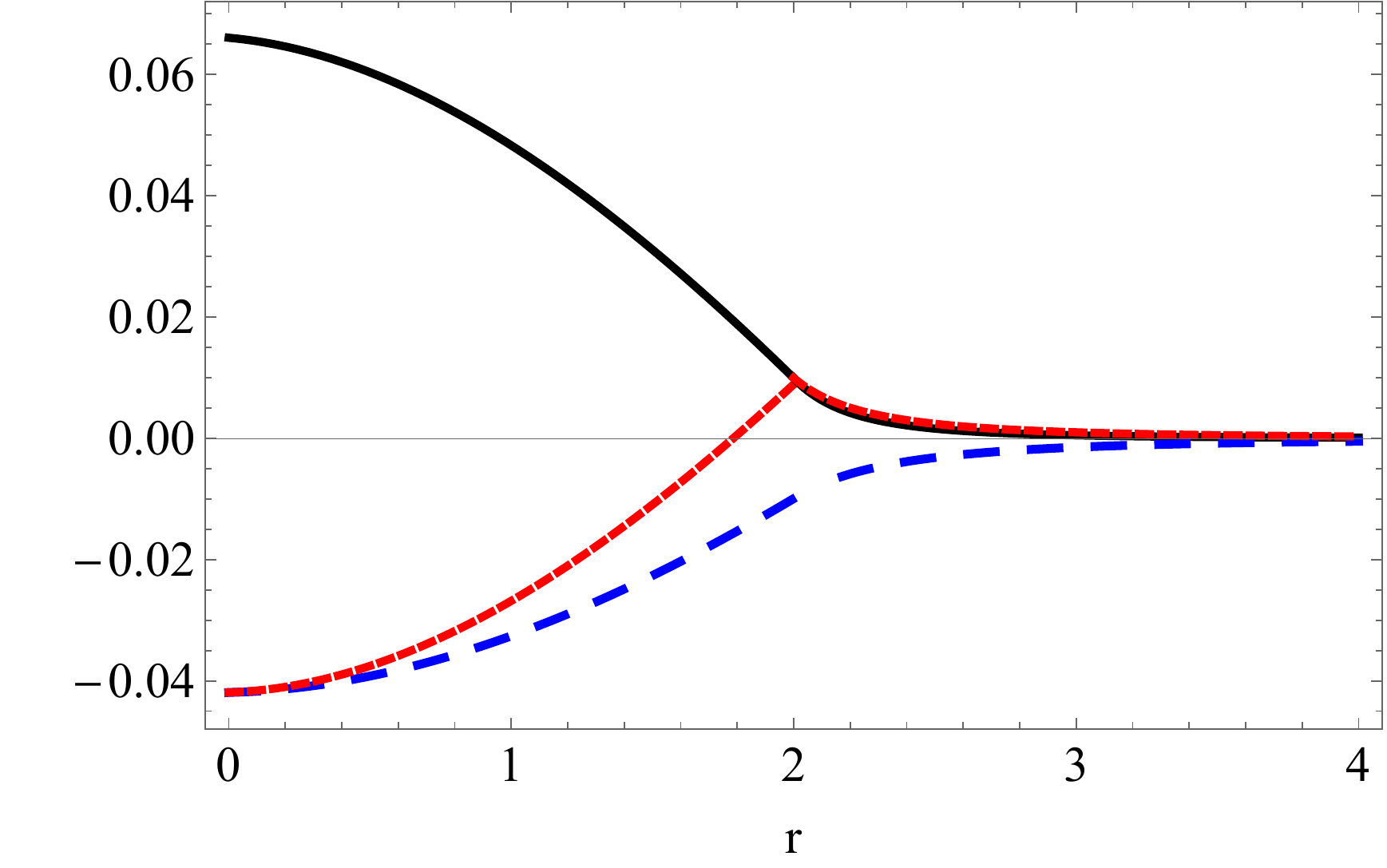}
\caption{\label{matter} $\rho$ (black line), $p_{r}$ (blue dashed line) and $p_{\perp}$ (red dotted line) for $M=1$ $a_{3}=0.01$ and $l=-1$ }
\end{figure}

In figure \ref{lambda} we show how the radial metric $e^{-\lambda}$ as a function of $r$ for the specific values in the legend.
Note that, as in Ref. \cite{ovalle2019a}, the metric function $e^{-\lambda}$ is smoothly continuous though the stellar surface. In figure \ref{matter} we show the matter sector for the specific values in the legend where we note that not only the radial pressure but the density are continuous though the surface $\Sigma$ in accordance with the results previously reported in \cite{ovalle2019a}. In addition, note that the cusp--like matching of the tangential defines the surface of the star.

\subsection{Exterior 2}
In this case we consider the MGD--deformed vacuum (see Ref. \cite{ovalle2018a} for details)
\begin{eqnarray}
e^{\nu}&=&1-\frac{2M}{r}\\
e^{-\lambda}&=&\left(1-\frac{2M}{r}\right)\left(1+\frac{\beta}{(r+M)^{2(a-1)}}\right)\\
\rho&=&\frac{\beta  (M+r)^{1-2 a} ((3-4 a) M+(2 a-3) r)}{8 \pi  r^2}\\
p_{r}&=&\frac{\beta  (M+r)^{2-2 a}}{8 \pi  r^2}\\
p_{\perp}&=&\beta\frac{(a-1)   (M-r) (M+r)^{1-2 a}}{8 \pi  r^2},
\end{eqnarray}
where $a>1$ to ensure asymptotic flatness and $\beta$ is the decoupling parameter of the exterior geometry. The continuity of the radial pressure leads to
\begin{eqnarray}
a_{2}&=&-\frac{1}{32} 3^{-2 a-1} M^{-2 a-4} (128\ 3^{2 a} a_{3} M^{2 a+5}\nonumber\\
&+&3^{2 a+2} M^{2 a}+27 \beta  M^2),
\end{eqnarray}
and, as in the previous case the continuity of the first fundamental form is satisfied by construction.\\
 
\begin{figure}[h!]
\includegraphics[scale=0.5]{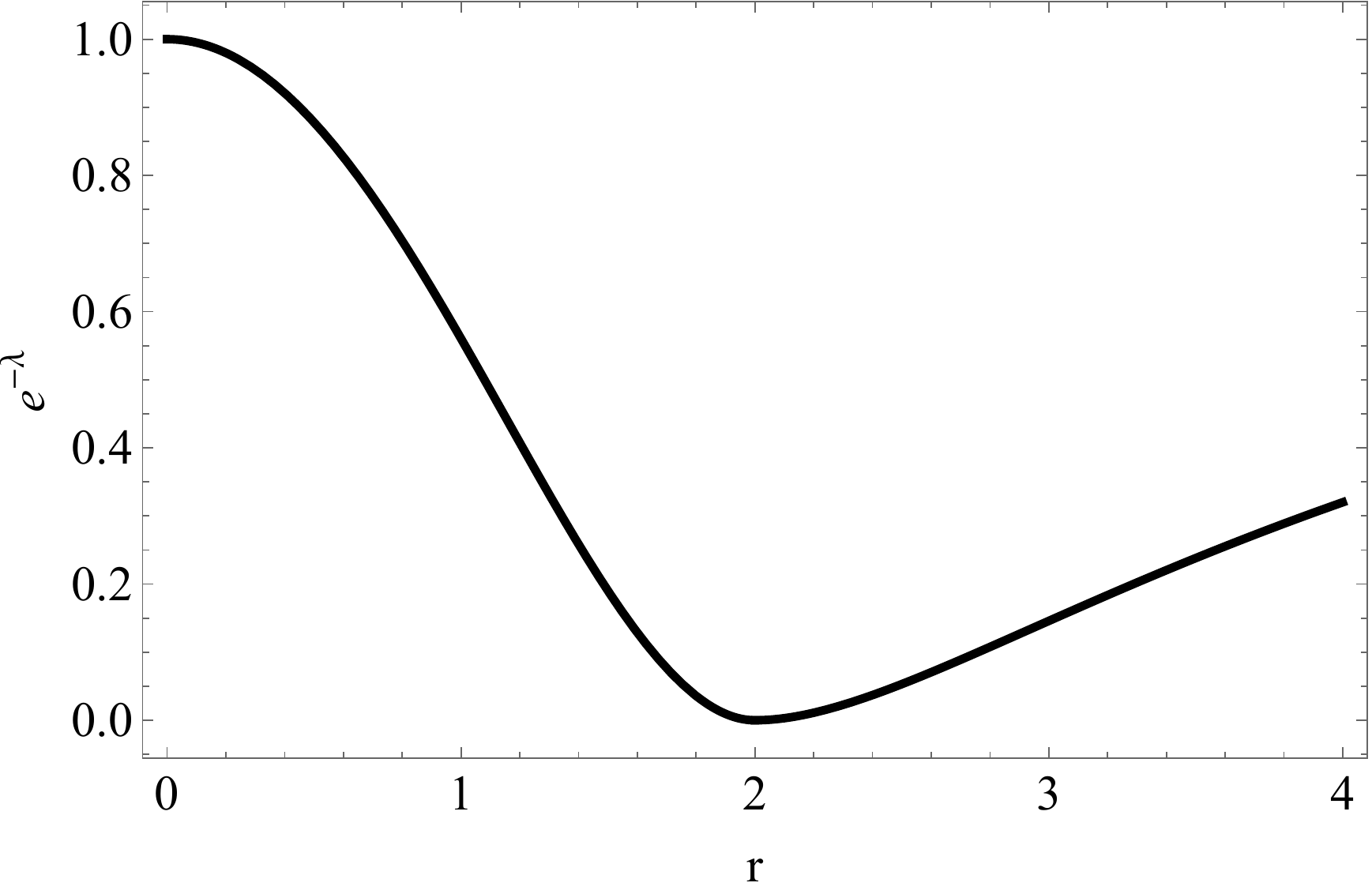}
\caption{\label{lambda2} $e^{-\lambda}$ for $M=1$ $a_{3}=0.01$, $\beta=-9$ and $a=2$}
\end{figure}

\begin{figure}[h!]
\includegraphics[scale=0.5]{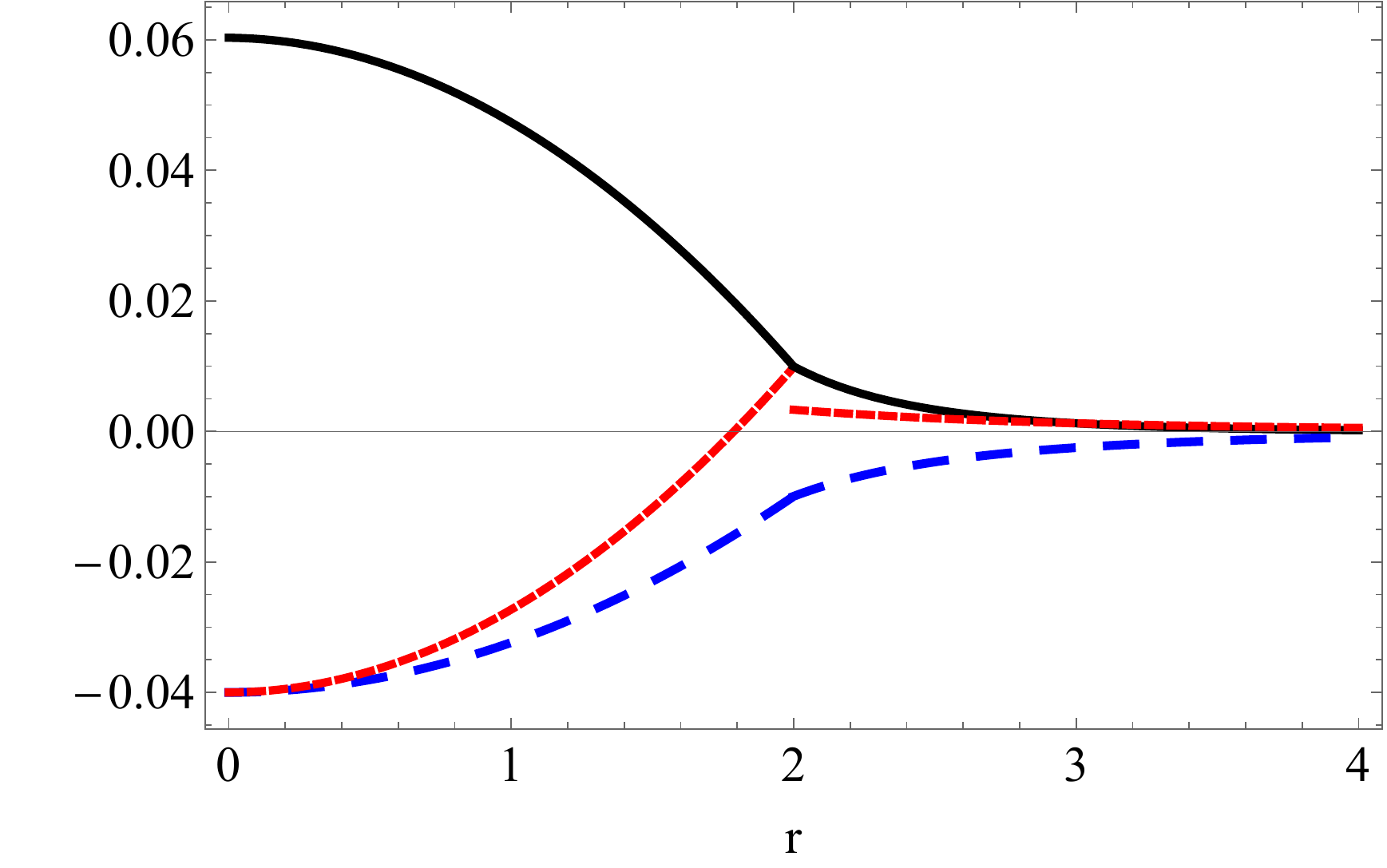}
\caption{\label{matter2}  $\rho$ (black line), $p_{r}$ (blue dashed line) and $p_{\perp}$ (red dotted line) for $M=1$ $a_{3}=0.01$, $\beta=-9$ and $a=2$}
\end{figure}

In figure \ref{lambda2} we show how the radial metric $e^{-\lambda}$ as a function of $r$ for the specific values in the legend. Again, it is noticeable the smooth behaviour of the metric function.
In figure \ref{matter2} we show the matter sector as a function of $r$ for the specific values in the legend and again, we note the continuity in both the radial pressure and the density. However, in contrast to the previous case, the tangential pressure is discontinuous at the surface.

\section{Conclusions}
In this work we constructed a new ultracompact aniso-\nobreak tropic star solution in the framework of the Gravitational Decoupling by the Minimal Geometric Deformation approach. As the auxiliary condition to close the system of differential equations we used the complexity factor of self--gravitating fluids. Inspired by the results found in 
\cite{ovalle2019a}, we proposed a polynomial complexity and obtained that the interior solution obtain can be well--matched to two different modified vacuum. The solution obtained here, contains the reported in Ref. \cite{ovalle2019a} as an special case. Ore findings here indicate that the solution fulfill the requirements of a stable configuration, namely, i) the solution is regular at the origin, ii) the mass and the radius are well defined, iii) the density is positive everywhere and decreases monotonically to the surface and iv) the radial pressure is non--uniform and monotonic as expected.\\

Although we only analysed the case $N=3$ here, it can be easily shown that higher orders can also provide suitable gravastar models for particular values of the free parameters involved. However,
it should be interesting to consider higher orders in the polynomial complexity matched to different modified vacuum to explore to what extend the model leads to well behaved solutions.

\end{document}